\documentstyle[epsf,aps,prb]{revtex}

\begin{document}

\title{Stability of homogeneous magnetic phases in a generalized $t-J$
model}

\author{ L. O. Manuel and H. A. Ceccatto}
\address {Instituto de F\'{\i}sica Rosario, Consejo Nacional de
Investigaciones Cient\'{\i}ficas y T\'ecnicas \\ and Universidad Nacional de
Rosario, Bvd. 27 de Febrero 210 Bis, 2000 Rosario, Rep\'ublica
Argentina }
\maketitle

\begin{abstract}
We study the stability of homogeneous magnetic phases in a
generalized $t-J$ model including a same-sublattice hopping $t^{\prime }$
and nearest-neighbor repulsion $V$ by means of the slave fermion-Schwinger
boson representation of spin operators. At mean-field order we find, in
agreement with other authors, that the inclusion of further-neighbor hopping
and Coulomb repulsion makes the compressibility positive, thereby
stabilizing at this level the spiral and N\'{e}el orders against phase
separation. However, the consideration of Gaussian fluctuation of order
parameters around these mean-field solutions produces unstable modes in the
dynamical matrix for all relevant parameter values, leaving only reduced
stability regions for the N\'{e}el phase. We have computed the one-loop
corrections to the energy in these regions, and have also briefly considered
the effects of the correlated hopping term that is obtained in the reduction
from the Hubbard to the $t-J$ model.
\end{abstract}

\vskip 0.5cm

Two-dimensional antiferromagnets doped with mobile holes have been
extensively studied in the last decade in order to understand the physics of
superconducting cuprates. Most often these studies have been carried out by
considering the so called $t-J$ model,

\begin{equation}
H_{tJ}=-t{\sum_{ij}}c_{i\sigma }^{\dagger }c_{j\sigma }+\frac{J}{2%
}{\sum_{ij}}{\bf S}_{i}\cdot {\bf S}_{j},  \label{HtJ}
\end{equation}
where $i,j$ are nearest-neighbor sites on a square lattice, the operator $%
c_{i\sigma }^{\dagger }$ creates a particle on a vacuum where all sites
carry a Zhang-Rice singlet, and double occupancy is forbidden ($c_{i\sigma
}^{\dagger }n_{i}=0$). When this Hamiltonian is derived from a realistic
three-band Hubbard model for the CuO$_{2}$ planes, the hopping integral $%
t\simeq 2.5$ in units of the coupling $J=1.$ At half filling (\ref{HtJ})
reduces to the Heisenberg antiferromagnet, which displays quantum N\'{e}el
order at $T=0$. The introduction of mobile holes disrupts the magnetic
order, and eventually melts it. The description of this quantum phase
transition has lead to the consideration of two related questions: i) the
existence of an intermediate spiral magnetic phase for small doping, and ii)
the possibility of phase separation in the ground state of the $t-J$ model .
The existence of a spiral phase is very appealing in view of the
incommensurate peaks in the magnetic susceptibility found in real materials;%
\cite{cheong91} this incommensurability seems to appear only in
superconducting samples, which lends support to the idea that is related to
hole mobility. Phase separation is possible since for $J<<t$ the $t-J$ model
is related to the large repulsion limit of the Hubbard model, which only
accounts for local on-site interactions. Consideration of the true
long-range Coulomb interaction, even if strongly screened, could then
prevent phase separation.

Shraiman and Siggia\cite{shraiman89} were the first to discuss the existence
of a spiral phase in the $t-J$ model. In a semiclassical approach, they
showed that for infinitesimal doping $\delta =1-n$, where $n$ is the
particle density, the lowest-energy spin configuration is a spiral with
pitch $Q\approx \delta t/J$. Since then, many authors have considered this
problem using a variety of methods. One of the most elegant approaches is
the slave fermion-Schwinger boson (SF-SB) theory, in which the physical
particle is replaced by a product of a spinless fermion $f$ and a bosonic
spinor ${\bf b^{\dagger }}=(b_{\uparrow }^{\dagger },b_{\downarrow
}^{\dagger })$: 
\begin{equation}
c_{i\sigma }\rightarrow b_{i\sigma }f_{i}^{\dagger },\qquad ~c_{i\sigma
}^{\dagger }\rightarrow f_{i}b_{i\sigma }^{\dagger }.  \label{rep}
\end{equation}
This replacement, together with the site-occupation constraint $\sum_{\sigma
}b_{i\sigma }^{\dagger }b_{i\sigma }+f_{i}^{\dagger }f_{i}=1$, is a faithful
representation of the original Fermi algebra. Thus, the spins can be written
in terms of the Schwinger bosons as ${\bf {S}}_{i}=\frac{1}{2}b_{i\sigma
}^{\dagger }{\bf {\sigma }}_{\sigma \sigma ^{\prime }}b_{i\sigma ^{\prime }}$
(${\bf {\sigma }}$ is the vector of Pauli matrices), while the slave fermion 
$f_{i}$ accounts for the charge degrees of freedom . This representation
naturally enforces the $t-J$ model restriction to a single particle per
site. Moreover, at half filling it correctly reproduces the spin-wave
results for the magnetization and spectrum, unlike the alternative fermionic
description of spins.\cite{affleck88} Using the representation (\ref{rep}),
Jayaprakash, Krishnamurthy and Sarker\cite{jayaprakash89} found a $T=0$
phase diagram with four phases: a metallic phase with long-range
ferromagnetic order for $t\delta >>J$, an insulating phase with
antiferromagnetic correlations for $t<<J$ and $\delta $ not too large, a
disordered ''spin-liquid '' insulating phase with only short-range
antiferromagnetic correlations for $t<<J$ and $\delta $ large, and a spiral
metallic phase with incommensurate spin order when $t\delta \simeq J$. 
About the same time, Kane {\it et al.}\cite{kane90} argued that the
incoherent processes that renormalize the hole spectrum in an
antiferromagnetic background may be treated by including an effective
same-sublattice hopping term $t^{\prime }.$ They considered several
homogeneous spin configurations (N\'{e}el,canted, $(Q,Q)$ and $(Q,\pi )$
spirals, double spiral) within the same SF-SB mean-field theory
of \cite{jayaprakash89}.
For small doping ($\delta \approx 0$) these authors found
that the spiral and N\'{e}el orders are favored, with their relative
stability depending on the value of $t^{\prime }.$ They also found that for
both $\delta $ and $t/J$ large a state with uniform chiral order is favored.
This state is similar to the flux phase obtained near half filling in the
alternative slave-boson approach, and violates parity and time-reversal
symmetries.

Antiferromagnetic systems are highly susceptible to local deformations of
the spin order in the presence of holes. Their homogeneous magnetic states
are thus inherently unstable due to the large compressibility of the holes,
since the system wishes to enhance the spin deviations by locally increasing
the hole density. This tendency to phase separation in the $t-J$ model was
observed by Auerbach and Larson\cite{auerbach91} (AL) by studying the SF-SB
mean-field states by means of a expansion in $\delta $. At the mean-field
level, the most important result of \cite{auerbach91}is the negativity of
the compressibility ${\kappa }$ for all values of $t/J$ and very small $%
\delta $. This negative ${\kappa }$ suggests that phase separation between a
hole-rich and a N\'{e}el phase takes place, provided the energy is minimized
by homogeneous states at finite doping. Alternatively, the system could form
local defects or spin polarons.\cite{auerbach91b} Other works using
different techniques lead to similar results.\cite
{ivanov91,igarashi92,mori93}  However, Mori and Hamada\cite{mori93} showed
that the inclusion of the $t^{\prime }$ term can stabilize a N\'{e}el phase
for small $\delta $. Furthermore, Sarker\cite{sarker93} found that in the
presence of a Coulomb interaction (screened to a simple nearest-neighbor
repulsion $V$) the incommensurate spiral states and also the
antiferromagnetic order are favored over the flux states of \cite{kane90},
and, more importantly, that they are stable against hole condensation for $%
\delta $ larger than some $\delta _{c}$.

Except for \cite{auerbach91}, all the works mentioned above using the SF-SB
representation are mean-field in character. The subtle differences among
them correspond only to the several ways that the exact form of the
constraint is used to simplify terms in the Hamiltonian. Since the
constraint is later relaxed, all this equivalent Hamiltonians lead to
slightly different results and, as a consequence, the relative stability of
the magnetic phases changes. The more interesting question of the general
stability of the mean-field picture due to order-parameter fluctuations was
partially considered by AL, who found that for
infinitesimal $\delta $ the spiral phases are unstable for all values of $t/J
$. This instability appears as a negative eigenvalue of the dynamical matrix
---the Hessian of the ground-state energy as a function of the order
parameters---. They also considered the effects of the intrasublattice
correlated hopping term of $O(t^{2}/U)$ that comes from the large $U/t$
transformation of the Hubbard model ---which was taking into account by the $%
t^{\prime }$ hopping term of lower symmetry in \cite{kane90}---, and
stressed the importance of this term in allowing the instability of the
magnetic states at infinitesimal doping.

Despite the above reviewed widespread use in the literature of the SF-SB
approach, we are not aware of a thorough investigation of fluctuations
effects on mean-field results. In particular, exactly at half filling ($%
\delta =0$) the general SF-SB theory reduces to the Schwinger-boson large-$N$
expansion for the Heisenberg antiferromagnet, recently extended by us\cite
{trumper97} to include the $O(1/N)$ corrections. In view of the results of
Sarker,\cite{sarker93} we believe it is worth reconsidering the stability of
homogeneous magnetic phases in the $t-$ $J-V$ model since, as stated above,
the conclusions of AL are restricted to infinitesimal $\delta $ and $V=0.$
In the following we will investigate this question by the consideration of
order-parameter fluctuations. In particular, the novel features introduced
in our calculations are the following: i) since the theory presents a local $%
U(1)$ gauge symmetry we will use collective coordinate methods ---as
developed in the context of relativistic lattice gauge theories\cite
{polyakov77}--- to handle the infinitely-many zero modes associated to the
local symmetry breaking in the saddle-point expansion. Contrary to what
happens in the radial gauge, in this case the dynamical matrix explicitly
displays the zero modes, which allows to check the calculations at
intermediate steps; ii) unlike in \cite{auerbach91}, here we will consider
finite densities and will evaluate the fluctuations corrections in the
region where the magnetic phases are stable; iii) we will include a
nearest-neighbor repulsion $V$ as considered by Sarker,\cite{sarker93} and
also a second neighbor hopping $t^{\prime }$ in the spirit of \cite{kane90}.
We stress that the presence of a same-sublattice hopping has been frequently
advocated as a stabilizing mechanism for spiral phases.\cite
{kane90,mori93,normand95} The correlated hopping term considered by AL
will be initially disregarded; its effects will be briefly
discussed at the end of this work. We emphasize that these calculations are,
to the best of our knowledge, the first {\it complete} evaluation of
fluctuations corrections in the SF-SB theory reported in the literature.

After the Hamiltonian is written in terms of the new operators (\ref{rep}),
the partition function can be expressed as a functional integral over
coherent Bose and Fermi states. Then, the different terms in ({\ref{HtJ})
are decoupled by means }of Hubbard-Stratonovich transformations according to
the following scheme (indicated by a vertical bar):

\[
H_{J}\longrightarrow \frac{J}{4}\sum_{ij}\left( :\hat{B}_{ij}^{\dagger }\ |\ 
\hat{B}_{ij}:-\hat{A}_{ij}^{\dagger }\ |\ \hat{A}_{ij}\right) 
\]
\begin{equation}
H_{t}\longrightarrow -\sum_{ij\ }t_{ij}\text{ }f_{i}f_{j}^{\dagger }\ |\ 
\hat{B}_{ij}  \label{decomp}
\end{equation}
\[
H_{V}\longrightarrow V\sum_{ij\ }f_{i}^{\dagger }f_{i\text{ \ }}|\text{ \ }%
f_{j}^{\dagger }f_{j}.
\]
In $H_{J\text{ }}$we wrote the Schwinger-boson spin-spin interaction in
terms of the two $SU(2)$ invariants $\hat{B}_{ij}^{\dagger }=\sum_{\sigma
}b_{i\sigma }^{\dagger }b_{j\sigma }$ and $\hat{A}_{ij}=\sum_{\sigma }\sigma
b_{i\sigma }b_{j\bar{\sigma}}$. These operators describe the magnetic
fluctuations in the ferromagnetic and antiferromagnetic channels
respectively, and allow for a proper treatment of the incommensurate order
upon doping. For the hopping term $H_{t}$ we just decoupled bosons from
fermions (here the indices $i,j$ run also between next-nearest neighbors).
Finally, for $H_{V}$ we indicated the Hartree factorization but we have also
considered i) the alternative Fock decoupling, and ii) the contribution of
this term when the nearest-neighbor repulsion is written in terms of bosons, 
{\it i.e.}, $H_{V}=V\sum_{ij\ }\hat{B}_{ii\text{ }}^{\dagger }$ $|$ $\hat{B}%
_{jj\text{ }}^{\dagger }$; in all these cases the results are qualitatively
the same. According to the decompositions (\ref{decomp}) of the interaction
terms, in a saddle-point evaluation of the partition function the
Hubbard-Stratonovich fields become the following self-consistent order
parameters: 1) antiferromagnetic order: $A_{i-j}=\langle \sum_{\sigma
}\sigma b_{i\sigma }b_{j\sigma }\rangle $, 2) ferromagnetic order: $%
B_{i-j}=\langle \sum_{\sigma }b_{i\sigma }^{\dagger }b_{j\sigma }\rangle $,
3) hopping: $F_{i-j}=\langle f_{i}^{\dagger }f_{j}\rangle $, and 4) hole
density $N_{i}=\langle f_{i}^{\dagger }f_{i}\rangle $. The saddle-point
Hamiltonian, quadratic in boson and fermion operators, can be diagonalized
in the standard way. Here we only point out that in transforming to momentum
space the boson operators we have defined $b_{i\sigma }^{\dagger }=\frac{1}{%
\sqrt{N}}\sum_{{\bf k}}b_{{\bf k}\sigma }^{\dagger }e^{{\bf k}_{\sigma .}%
{\bf r}_{i}}$, with ${\bf k}_{\sigma }={\bf k}+\sigma {\bf Q}/2.$ The extra
phase $\sigma {\bf Q.r}/2$ is required to obtain the gapless (Goldstone)
modes associated to the long-range magnetic order at the proper ${\bf k}%
=0,\pm {\bf Q}$ points.\cite{ceccatto93} The self-consistent order
parameters are determined using a mesh of 100$\times $100 points for the
integrals in reciprocal space. Since our mean-field theory is not
variational because it unphysically enlarges the configuration space, we
cannot rely only on the energy-minimization criterion to seek for solutions
of these consistency equations. Consequently, based on previous studies of
the $t-J$ model, we considered solutions corresponding to magnetic
wavevectors $(\pi -Q,\pi )$ and $(\pi -Q,\pi -Q)$, which in large regions of
parameter space are very close in energy.

At the mean field level we computed the hole compressibility $\kappa \sim 
\frac{\partial ^{2}E_{{\rm GS}}}{\partial n^{2}}$ to investigate indications
of possible phase separation. In Fig. 1 we plot the energy of the $(Q,Q)$
spiral phase as a function of the density of holes for the standard $t-J$
model ($t^{\prime }=0=V$). From this figure we see that, as found in several
studies, the compressibility of holes is negative for small densities and
all values of $t/J$, though it becomes positive for large to moderate doping
and large $t/J$. The inclusion of a second-neighbor hopping $t^{\prime }$
makes the spirals and N\'{e}el phases to be even closer in energy (less than
1\% of difference) for all relevant parameter values. In Fig. 2 we give the
energy of the N\'{e}el phase as a function of doping for different values of 
$t^{\prime }/J$ and $V=0$. We found that this energy does not depend on $t$,
since at mean-field order in the N\'{e}el phase the dominant processes
correspond to coherent hopping in the same sublattice. Fig. 2 shows that for 
$t^{\prime }/J>0.05$ the compressibility becomes positive (the same is valid
for the spiral phases and holds in general for $\vert t^{\prime
}|/J>0.05$): the non-frustrating character of the second-neighbor hopping
promotes hole mobility and prevents their segregation in real space. We have
also verified that the inclusion of the nearest-neighbor repulsion $V$
further favors the stability of the homogeneous magnetic phases, as found by 
Sarker. \cite{sarker93} Then, possible instabilities of these phases beyond
mean-field order are not necessarily related to phase separation.

We turn now to the consideration of Gaussian fluctuation corrections to
mean-field results, which can be obtained by extending the calculations
described in \cite{trumper97}. The saddle-point expansion breaks the gauge
invariance of the theory, leading to the existence of saddle-point solutions
connected by the continuous $U(1)$ gauge group. This makes the dynamical
matrix ${\cal D}$ to have infinitely-many zero modes, which are the
Goldstone bosons associated to the (spurious) local symmetry breaking. In
particular, for translational-invariant saddle-point values of the
decoupling fields, there is a zero mode ${\vec{\varphi}}_{0}^{R}({\bf k}%
,\omega )\!=\!{\delta {\vec{\varphi}}({\bf \theta })/{\delta \theta _{{\bf k}%
\omega }|}}_{\theta =0}{\ }$ in every ${\bf k}\!-\!\omega $ subspace. Here ${%
\vec{\varphi}}({\bf \theta })$ is the vector of gauge-transformed fields,
which depends on the set of local phases ${\bf \theta }${\bf ,} and the ${%
\vec{\varphi}}_{0}^{R}$'s can be shown to be {\it right} eigenvectors of the
nonhermitian matrix ${\cal D}$. To avoid the infinities associated to these
modes without restoring forces we introduce collective coordinates along the
gauge orbit. Exact integration of these coordinates eliminates the zero
modes and restore the gauge symmetry (in the sense that noninvariant
operators average to zero).\cite{brezin82} This program can be carried out by
enforcing in the functional integral for the partition function the
so-called background gauge condition or ``natural'' gauge, \cite{polyakov77}
${\vec{%
\varphi}}({\bf \theta })^{\dagger }.{\vec{\varphi}}_{0}^{R}=0$. This
condition can be introduced into the functional measure by means of the
Faddeev-Popoff trick, and restricts the integration to fields fluctuations
which are orthogonal to the collective coordinates. At $T=0$, after carrying
out these integrations on the genuine fluctuations, we obtain the one-loop
correction to the ground-state energy {\it per site}, 
\begin{equation}
E_{1}=-\frac{1}{2\pi }\int_{-\infty }^{\infty }d\omega \sum_{{\bf k}}\ln
\left( \frac{\Delta _{{\rm FP}}({\bf k},\omega )}{|\omega |\sqrt{\det {\cal D%
}_{\perp }({\bf k},\omega )}}\right) .  \label{E1}
\end{equation}
Here ${\cal D}_{\perp }$ is the projection of ${\cal D}$ in the subspace
orthogonal to the collective coordinates, and $\Delta _{{\rm FP}}({{\bf k}%
,\omega })\!=\!\left| {\vec{\varphi}}_{0}^{L}({{\bf k},\omega }).{\vec{%
\varphi}}_{0}^{R}({{\bf k},\omega })\right| $ is the Fadeev-Popoff
determinant (${\vec{\varphi}}_{0}^{L}({{\bf k},\omega })$ is the {\it left}
zero mode of ${\cal D}$ in the ${\bf k}\!-\!\omega $ subspace). More
explicitly, $\Delta _{{\rm FP}}({{\bf k},\omega })=\left[ 4\sum_{{\bf r}%
}J(1+\cos {\bf k.r})A_{{\bf r}}^{2}-4\sum_{{\bf r}}(JB_{{\bf r}}^{2}+4t_{%
{\bf r}}B_{{\bf r}}F_{{\bf r}})(1-\cos {\bf k.r})+\omega ^{2}\right] ^{1/2}$.

For spiral phases ${\cal D}_{\perp }$ has always negative eigenvalues, which
make these phases unstable for all parameter values even in the presence of
a large repulsion $V$. Only the N\'{e}el phase becomes stable in some
regions of the $(\delta ,t^{\prime }/J)$ plane because of the
intrasublattice hopping term, as shown in Fig. 3 for different values of $t/J
$ and $V=0$. Unlike the mean-field results, in this case the energy depends
on $t/J$ since the fluctuations corrections incorporate the incoherent
intersublattice hopping processes. Fig. 4 shows the corresponding stability
diagram for $t/J=2.5$, which is the value supposed to be relevant for the
cuprates. The consideration of both positive and negative values of $%
t^{\prime }$ is interesting in view of the proposal of Tohyama and Maekawa%
\cite{tohyama94} to explain the asymmetry in the behavior of superconducting
cuprates doped with holes or electrons. According to this proposal, the
asymmetry is related to the fact that in the generalized $t-J$ model for the
CuO$_{2}$ planes one should have $t^{\prime }>0$ for systems doped with
electrons and $t^{\prime }<0$ for those doped with holes. The results in
Fig. 4 are qualitatively supporting this idea, since the magnetic order in
systems with $t^{\prime }>0$ (corresponding to electron doping) are more
stable than those with $t^{\prime }<0$ (hole doping), as observed
experimentally. Notice, however, that there is also a region of stability
for small negative $t^{\prime }$ and medium to large $\delta $, but this
could be an artifact of the approximation. This picture is not modified by
the presence of the nearest-neighbor repulsion $V$. For completeness, in
Fig. 5 we plot the mean-field and fluctuation-corrected 
energies in the N\'eel phase 
as a function of doping on the stable part of the line $t^{\prime }=J$.

We have identified the unstable modes in the dynamical matrix. These modes
have $\omega \simeq 0$ and are distributed over most of the Brillouin
zone, although they are more densely located near ${\bf k=0}$ and $(\pi ,\pi
)$. They couple the fluctuations of the intersublattice fermionic hopping $%
F_{{\bf k}}$ and ferromagnetic order parameter $B_{{\bf k}}$. This clearly
shows that the tendency of the kinetic term $H_{t}$ in (\ref{HtJ}) to
locally align the spins is responsible for the instabilities of the
homogeneous magnetic phases, as previously found by AL.\cite{auerbach91}
This effect of local distortions is very difficult to account for in
mean-field theories, and requires the calculations of fluctuations to treat
it. The inclusion of the $V$ term reduces the number of unstable modes near $%
{\bf k=0\ }$(as expected, since this term acts against phase separation) but
it is not able to stabilize the homogeneous phases, a result that had also
been conjectured in \cite{auerbach91}.

Finally, we comment here on the effects of including the three-site
correlated hopping term of $O(t^{2}/U)$ that can be obtained in the
reduction of the Hubbard model to the $t-J$ one.\cite{auerbach91} After the
replacement (\ref{rep}) this term becomes a product of eight boson and
fermion operators, requiring a double Hubbard-Stratonovich transformations
to be decoupled. Using the same order parameters as in (\ref{decomp}), we
found that its inclusion does not modify qualitatively the above results. In
particular, it does not play any special roll in stabilizing or
destabilizing the magnetic phases. In a previous work\cite{batista97} we
considered the alternative Bogoliubov factorization of the four
slave-fermion operators in this term, which at mean-field order lead to a
superconducting phase with some interesting properties, similar to those
observed experimentally. Unfortunatelly, we found here that when the
order-parameter fluctuations are considered this superconducting phase
becomes unstable like the spiral phases in the magnetic state.

In this work we have considered, within the SF-SB theory, the general
stability of homogeneous magnetic phases in a generalized $t-J$ model
including a same-sublattice hopping $t^{\prime }$ and nearest-neighbor
repulsion $V$. At the mean-field level we found, in accordance with previous
works, that the new terms help to stabilize the N\'{e}el and spiral phases
against phase separation, since for small values of these parameters the
compressibility $\kappa $ becomes positive (see Fig. 2). The correlated
hopping term considered in \cite{auerbach91} does not substantially modify
these results. Beyond mean-field theory, the study of the dynamical matrix
associated to the order-parameter fluctuations reveals that the homogeneous
spiral phases are unstable in all regions of interest. On the other hand,
the N\'{e}el phase becomes stable for reasonable values of $t^{\prime }>0$
or unphysically large $t^{\prime }<0$, as shown in Figs. 3 and 4. The
consideration of moderate values of the nearest-neighbor repulsion $V$ does
not modify these figures. We stress that Fig. 4 supports qualitatively the
proposal in \cite{tohyama94} that links the asymmetric behavior of
hole-doped and electron-doped cuprates to the $t^{\prime }$ sign in the
corresponding effective $t-J$ model for the CuO$_{2}$ planes. Finally, we
found that the consideration of Gaussian corrections to the mean-field
picture of \cite{batista97} unfortunately destroys the stability of the
superconducting phase there obtained.

\acknowledgements 
{We acknowledge useful discussions with A. A.
Aligia and C. D. Batista from Centro At\'{o}mico Bariloche.}

\figure{Fig. 1:  Mean-field energy $E(Q,Q)$ 
of the $(Q,Q)$ spiral phase
as a function of doping $\delta$ 
for different values of $t/J$ and $t^{\prime }=0=V$.}

\figure{Fig. 2:  Mean-field energy $E(\pi,\pi)$ 
of the N\'{e}el phase as a
function of doping $\delta$ 
for different values of $t^{\prime }/J$ and $V=0$.} 

\figure{Fig. 3:  Stability of the N\'{e}el phase for $t^{\prime }>0$, $V=0$%
, and different values of $t/J$. The antiferromagnetic 
order is stable in the regions 
above the lines.  } 

\figure{Fig. 4:  Stability diagram of the N\'{e}el phase in the whole $%
(\delta ,t^{\prime }/J)$ plane for the physically relevant value $t/J=2.5$.}

\figure{Fig. 5:  Mean-field  and fluctuation-corrected energies 
in the N\'{e}el phase as a function of doping $\delta$ 
for the parameter values $%
t/J=2.5$, $t^{\prime }=J$, and $V=0$, $J$.}

\newpage

\begin{figure}[ht]
\begin{center}
\epsfysize=10cm
\leavevmode
\epsffile{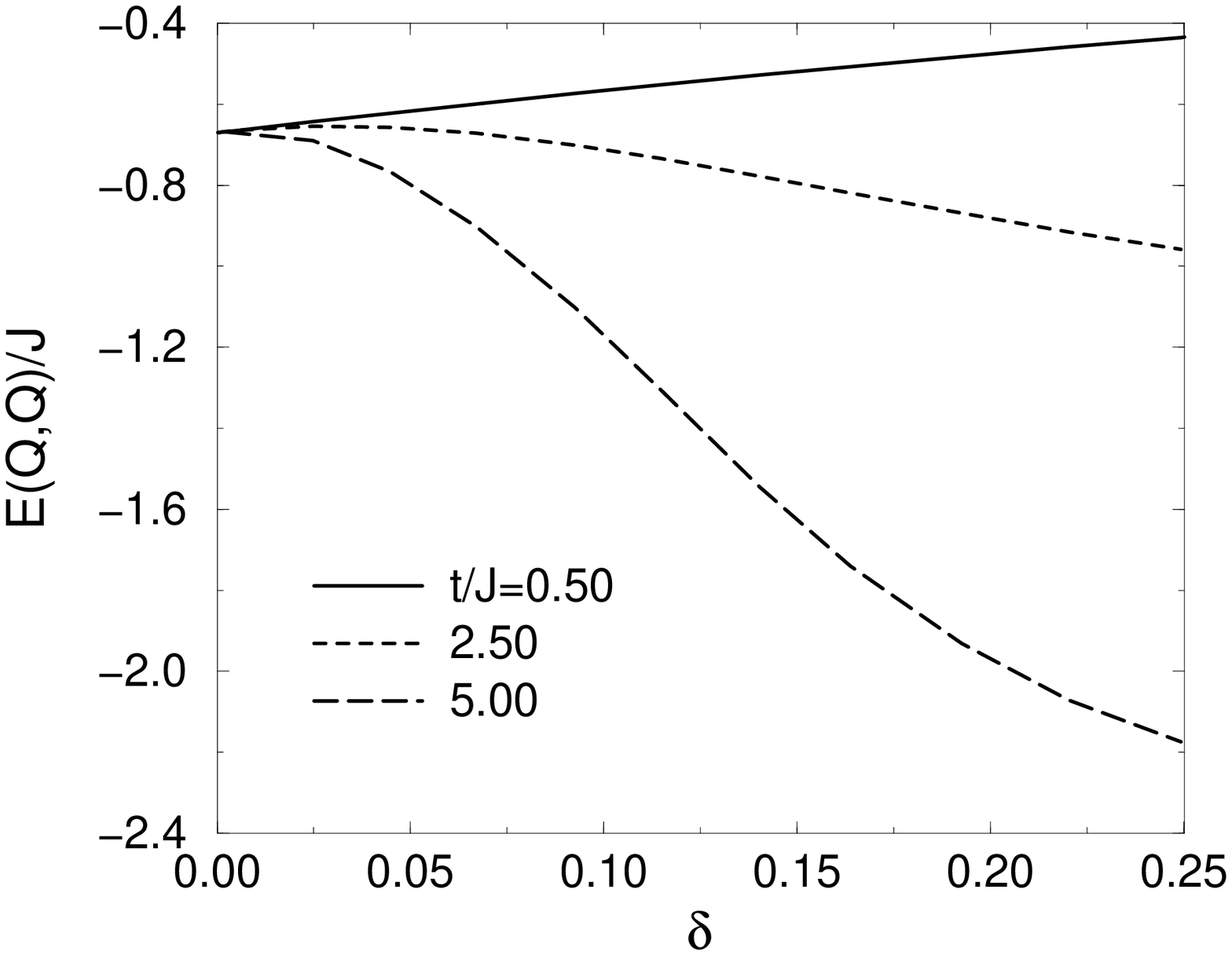}
\caption{}
\end{center}
\end{figure}

\begin{figure}[ht]
\begin{center}
\epsfysize=10cm
\leavevmode
\epsffile{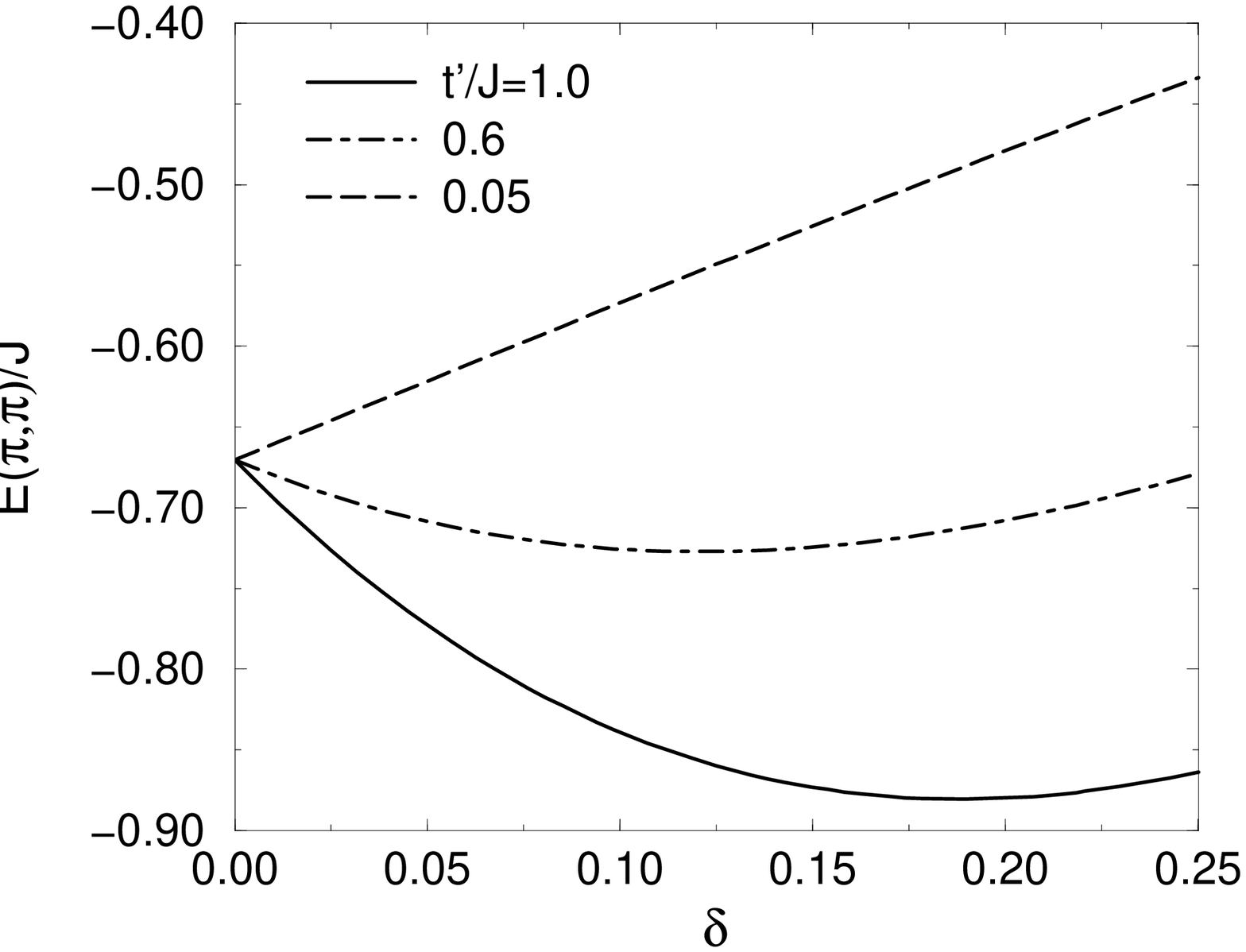}
\caption{}
\end{center}
\end{figure}

\begin{figure}[ht]
\begin{center}
\epsfysize=10cm
\leavevmode
\epsffile{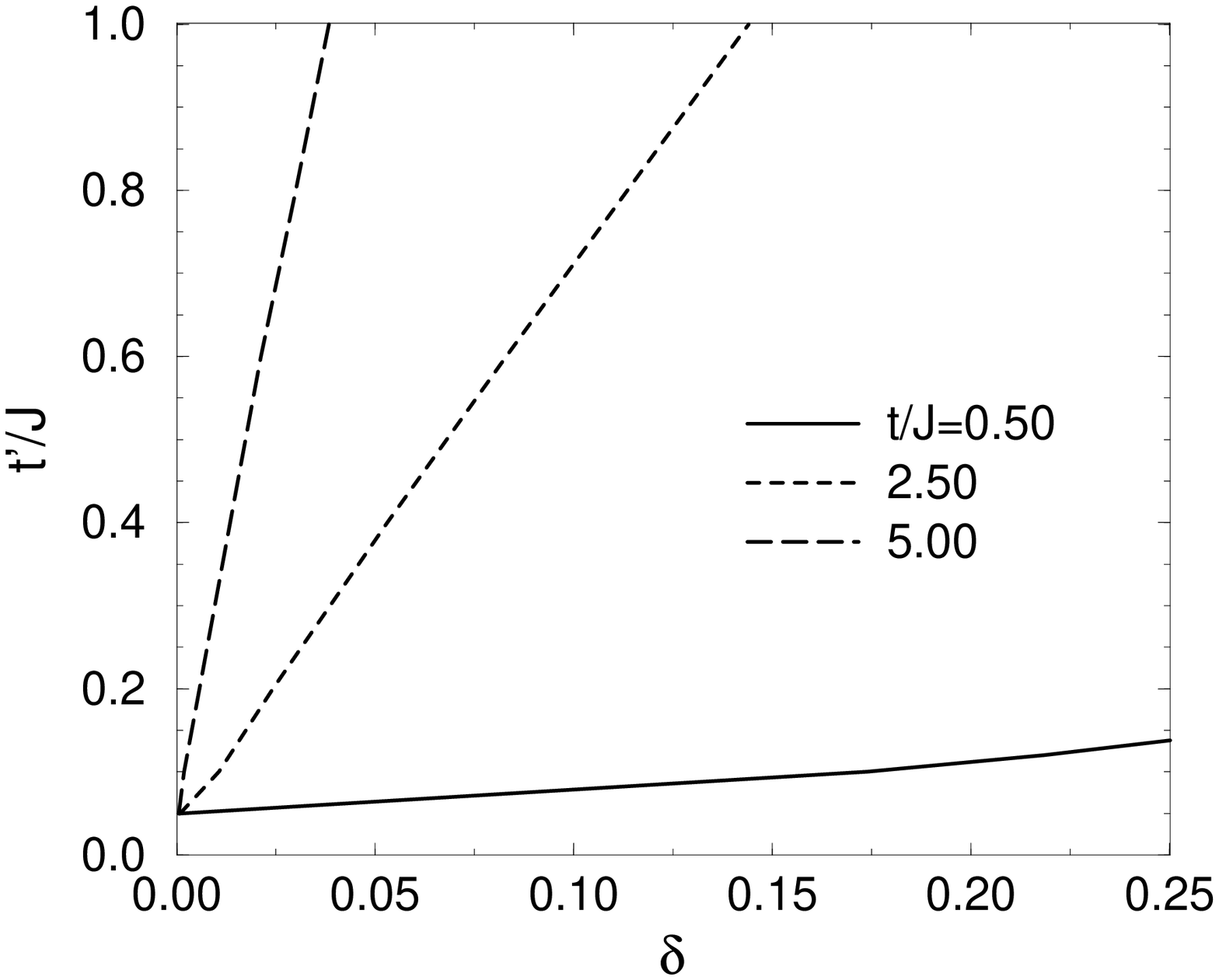}
\caption{}
\end{center}
\end{figure}

\begin{figure}[ht]
\begin{center}
\epsfysize=10cm
\leavevmode
\epsffile{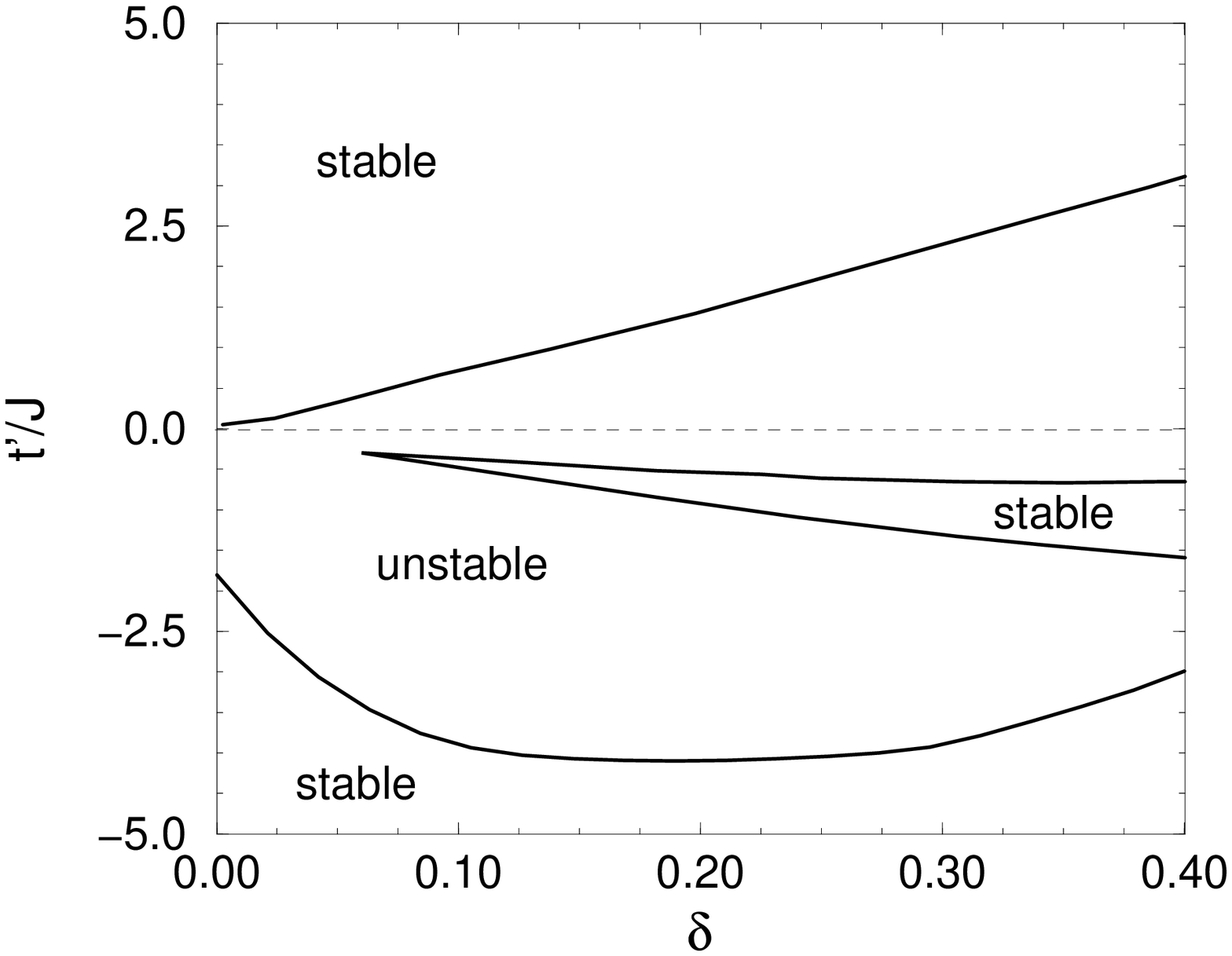}
\caption{}
\end{center}
\end{figure}

\begin{figure}[ht]
\begin{center}
\epsfysize=10cm
\leavevmode
\epsffile{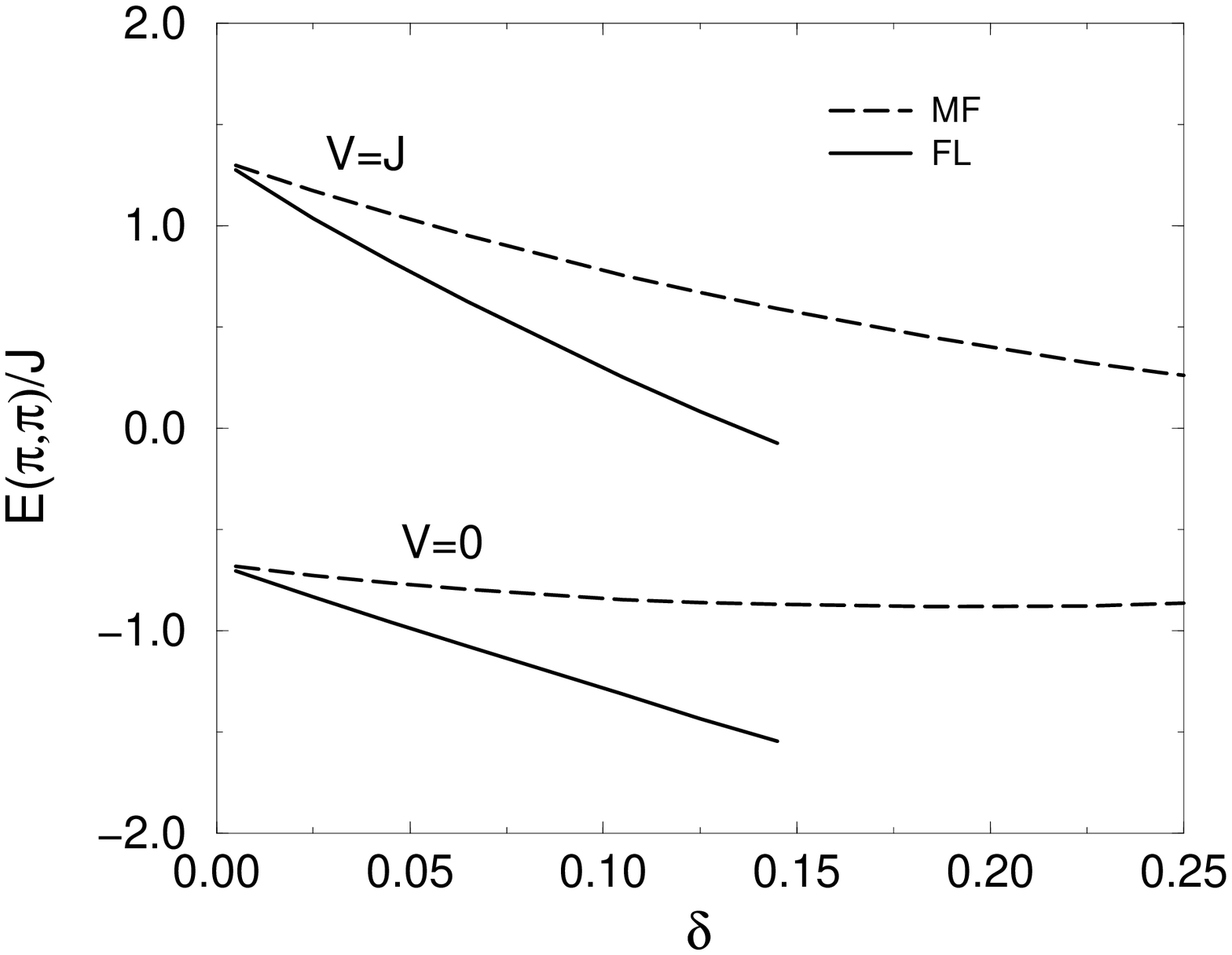}
\caption{}
\end{center}
\end{figure}

\end{document}